# Epitaxial Growth and Anomalous Hall Effect in High-Quality Altermagnetic α-MnTe Thin Films

Tian-Hao Shao[1,#], Xingze Dai[1,#], Wenyu Hu[1], Ming-Yuan Zhu[1], Yuanqiang He[2], Lin-He Yang[1], Jingjing Liu[1], Meng Yang[1], Xiang-Rui Liu[1], Jing-Jing Shi[1], Tian-Yi Xiao[1], Yu-Jie Hao[1], Xiao-Ming Ma[1,3], Yue Dai[1], Meng Zeng[1], Qinwu Gao[1], Gan Wang[1], Junxue Li[1], Chao Wang[2], and Chang Liu[1,*]

[1]*Department of Physics, State Key Laboratory of Quantum Functional Materials, and Guangdong Basic Research Center of Excellence for Quantum Science, Southern University of Science and Technology (SUSTech), Shenzhen, Guangdong 518055, China*

[2]*Institute of Nanosurface Science and Engineering, College of Mechatronics and Control Engineering, Shenzhen University, Shenzhen 518060, China*

[3]*College of Integrated Circuits and Optoelectronic Chips, Shenzhen Technology University, Shenzhen 518118, Guangdong, China*

\# These authors contributed equally: Tian-Hao Shao, Xingze Dai

*Corresponding author: Chang Liu

Email: liuc@sustech.edu.cn

## Abstract

The recent identification of α-MnTe as a candidate altermagnet has attracted considerable interest, particularly for its potential application in magnetic random-access memory. However, the development of high-quality thin films — essential for practical implementation — has remained limited. Here, we report the epitaxial growth of centimeter-scale α-MnTe thin films on InP(111) substrates via molecular beam epitaxy (MBE). Through X-ray diffraction (XRD) analysis, we construct a MnTe phase diagram that provides clear guidance for stabilizing the pure α-MnTe phase, revealing that it is favored under high Te/Mn flux ratios and elevated growth temperatures. Cross-sectional electron microscopy confirms an atomically sharp film–substrate interface, consistent with a layer-by-layer epitaxial growth mode. Remarkably, these high-quality α-MnTe films exhibit a pronounced anomalous Hall effect (AHE) originating from Berry curvature, despite a net magnetic moment approaching zero — a signature of robust altermagnetic character. Our work establishes a viable route for synthesizing wafer-scale α-MnTe thin films and highlights their promise for altermagnet-based spintronics and magnetic sensing.

# 1. Introduction

Spin-split antiferromagnetism (including altermagnetism) is a novel magnetic phase characterized by a non-relativistic spin-splitting arising from a compensated magnetic order (collinear for altermagnets), distinguishing it from conventional ferro- and antiferromagnets [1-6]. This unique nature enables these magnets to exhibit not only strong time-reversal-symmetry-breaking magneto-responses comparable to ferromagnets, but also key antiferromagnetic advantages including negligible stray fields, exceptional robustness against perturbations, and terahertz-frequency dynamics [2]. Representative spin-split antiferromagnetic materials include $MnTe_2$ [7], α-MnTe [8], CrSb [9], $Mn_5Si_3$ [10], Rb-doped $V_2Te_2O$ [11], and K-doped $V_2Se_2O$ [12], among others. In this work, we focus on α-MnTe owing to its high Néel temperature ($T_N$ > 300 K), pronounced anomalous Hall effect (AHE) and strong spin-splitting properties, which collectively establish it as an exceptional candidate for spintronic applications and next-generation magnetic random-access memory with low power consumption, high-speed operation, and high storage density.

α-MnTe, a room-temperature altermagnet, has recently been theoretically predicted and experimentally verified to exhibit strong momentum-dependent spin splitting [13-19], rendering it highly promising for applications in spin-transfer torque memories, spin-orbit torque devices, and magnetic sensors. However, the synthesis of high-quality, large-scale α-MnTe thin films—essential building blocks for practical devices—remains a critical challenge. To address this issue, several growth techniques have been explored, including chemical vapor deposition (CVD), pulsed laser deposition (PLD), magnetron sputtering, and molecular beam epitaxy (MBE). Although CVD enables the production of high-quality α-MnTe [20-22], the resulting crystal dimensions are generally confined to the micrometer scale, limiting its applicability in large-scale electronic integration. Meanwhile, MnTe films fabricated by PLD and magnetron sputtering typically exhibit polycrystalline structures [23-25], which may impair device performance. In contrast, the MBE method stands out as particularly suitable for the high-quality, large-area growth of MnTe films, benefiting from its ultra-high vacuum environment, wide growth temperature window, and compatibility with diverse substrates.

Recently, molecular beam epitaxy (MBE) has been utilized for the growth of α-MnTe thin films. For example, Jain et al. demonstrated the epitaxial growth of α-MnTe on $Al_2O_3$ substrates with different buffer layers. Their results revealed that phase-pure α-MnTe films were achieved using a $Cr_2O_3$-$Bi_2Te_3$ buffer, whereas a mixed phase comprising both zinc-blende (γ-MnTe) and wurtzite (β-MnTe) polymorphs formed when an $In_2Se_3$ buffer was employed [26]. These observations underscore the crucial role of the buffer layer in phase selection—a factor that must be carefully accounted for in studies targeting the intrinsic properties of MnTe. Similarly, Zhu et al. reported the growth of α-MnTe films on $SrTiO_3$ substrates, though their films consistently contained γ-MnTe inclusions [27]. Meanwhile, Bey et al. grew α-MnTe on GaAs (111) substrates and observed a significant lattice misfit at the interface, leading to substantial strain in the α-MnTe layer [28]. Chilcote et al. also grew α-MnTe films on InP (111); however, neutron diffraction revealed that Mn-rich defects are introduced during the growth process, inducing ferromagnetism [29]. Collectively, these studies indicate that achieving high-quality, phase-pure α-MnTe films remains challenging. Given that InP (111) offers the smallest lattice mismatch with α-MnTe, we have selected this substrate for MBE growth of α-MnTe films. The excellent lattice match of InP (111) has made it a preferred substrate for investigating intrinsic properties of α-MnTe, such as its spin-split band structure and the anomalous Hall effect.

In this study, high-quality α-MnTe thin films were epitaxially grown on InP (111) substrates by systematically varying the Te/Mn flux ratio and growth temperature. This approach enabled the construction of a phase diagram for

MnTe films as a function of flux ratio and growth temperature, which provides clear guidance for selectively stabilizing the pure α-MnTe phase. Structural analyses confirm an atomically sharp film–substrate interface and a layer-by-layer epitaxial growth mode. Moreover, magnetotransport measurements reveal a pronounced anomalous Hall effect in the absence of a net magnetic moment, thus confirming the successful growth of high-quality α-MnTe films. Our findings not only establish a reliable fabrication route for wafer-scale altermagnetic α-MnTe thin films but also showcase their outstanding structural and transport properties, thereby paving the way for the integration of altermagnets into next-generation spintronic devices.

## 2. Results and Discussion

Altermagnetic α-MnTe adopts a hexagonal NiAs-type structure, belonging to space group P6$_3$/*mmc* (No. 194), as depicted in Fig. 1a. The magnetic moments of Mn atoms are aligned parallel within each *c*-plane but antiparallel between adjacent *c*-planes, thereby forming two spin-opposite sublattices. These sublattices are interconnected via a nonsymmorphic symmetry operation consisting of a sixfold rotation $C_6$ combined with a half-unit-cell translation $t_{1/2}$ along the *c*-axis, i.e., a screw axis, as indicated by the black dashed line and black arrow in Fig. 1a, respectively. Fig. 1b illustrates the orientation of the Néel vector within the *c*-plane of α-MnTe. In the configuration, the Mn moments are aligned along the $(1\bar{1}00)$ magnetic easy axis. For this orientation, the relativistic magnetic point group *m'm'm* is generated by inversion $\mathcal{P}$, a twofold rotation $C_2$ about the *c*-axis, and a mirror plane perpendicular to the Néel vector combined with time-reversal symmetry $\mathcal{MT}$. This spin-space-group symmetry allows for an anomalous Hall effect (AHE) pseudovector to be oriented along the *c*-axis, which makes it possible to observe a hysteretic AHE signal. The corresponding details will be discussed in the following sections [30].

Fig. 1c depicts a schematic diagram of the first Brillouin zone and its high-symmetry points for α-MnTe. The corresponding electronic structure along the L–Γ–L′ high-symmetry direction, calculated by density functional theory (DFT), is presented in Fig. 1d. Governed by its spin-space-group symmetry, the spin degeneracy is lifted even without the inclusion of spin-orbit coupling (SOC) [31], revealing a spin-polarized band structure where red and blue represent the positive and negative spin states parallel to the Γ-M direction ($\pm S_y$), respectively — a hallmark feature of altermagnets. The electronic structures along other high-symmetry directions, as well as the spin-polarized band structures along the L–Γ–L′ direction, both calculated with SOC, are presented in Fig. S1 of the Supplementary Information.

Fig. 1e presents the X-ray diffraction (XRD) pattern of the α-MnTe film grown on an InP(111) substrate by MBE. The prominent (000$l$) diffraction peaks of the α-MnTe film are clearly identifiable despite the much stronger InP(111) substrate peak. No additional diffraction peaks are observed, indicating the successful growth of a phase-pure α-MnTe film. An enlarged view of the (0002) diffraction peak from the region enclosed by the red rectangle in Fig. 1e is shown in Fig. 1f, where the peak is clearly marked by a red arrow. Furthermore, the weaker fringes surrounding the (0002) diffraction peak, identified as Laue oscillations resulting from coherent diffraction at the top and bottom film interfaces, are also visible. These oscillations indicate that the crystalline interfaces are nearly atomically sharp. The left inset shows a sharp reflection high-energy electron diffraction (RHEED) pattern of the α-MnTe film, indicating a layer-by-layer growth mode. The right inset presents an optical microscope image, revealing that the films extend to the centimeter scale. This large-area, high-quality growth provides a solid foundation for subsequent device-oriented studies of α-MnTe.

To establish universal growth conditions for α-MnTe epitaxial films, a series of 30-nm-thick MnTe films were synthesized by systematically varying the Te/Mn flux ratio (4.865 – 14.635) and growth temperature (250 – 500°C). In all growth attempts, the Mn source temperature was kept constant at 750°C, while the Te/Mn flux ratio was adjusted by changing the temperature of the Te source. The calculation of the flux ratio, determined using a beam flux monitor, is detailed in Fig. S2. The crystal phases were confirmed by high-resolution XRD ω–2θ scans, selected data are shown in Fig. 2a-d. Specifically, Fig. 2a and b show the flux-dependent XRD patterns at fixed growth temperatures of 300°C and 500°C, respectively.

In Fig. 2a, the two sharp peaks with the highest intensity are attributed to the InP(111) substrate. The characteristic α-MnTe peaks (labeled @1 and @2) locate very close but to the right of the substrate peaks, as indicated by the grey dashed lines. Two additional peaks (labeled #1 and #2) to the left of the substrate peaks are also clearly observed (also marked by grey dashed lines), which are identified as originating from the zincblende phase of MnTe (γ-MnTe) [32]. The experiment indicates that, at a growth temperature of 300°C, the Te/Mn flux ratio is the key parameter determining the phase composition of the MnTe thin films. When the flux ratio is below 6.962, only the pure γ-MnTe phase is formed. Once the flux ratio exceeds this critical value, both the α-MnTe and γ-MnTe phases appear simultaneously in the films, resulting in a mixed-phase structure. When the growth temperature was increased to 500°C (Fig. 2b), a primary diffraction peak (#1) of the γ-MnTe phase was observed only at a flux ratio of 5.827. Under all other flux ratios tested at this temperature, the films consisted solely of the α-MnTe phase. Notably, additional diffraction peaks appeared at flux ratios of 4.865 and 5.827, which were experimentally identified as originating from elemental In and Te, respectively (For detailed experimental data, please refer to Fig. S3 and Fig. S4 in the Supplementary Information).

Similarly, the growth-temperature-dependent XRD patterns at fixed flux ratios of 4.865 and 14.635 are shown in Fig. 2c and 2d, respectively. At a flux ratio of 4.865, the films are pure γ-MnTe below 450°C and transition to pure α-MnTe above this temperature (Fig. 2c). In contrast, at the higher flux ratio of

14.635, the pure α-MnTe phase becomes stable at a significantly lower temperature of 350°C (Fig. 2d). Notably, a polycrystalline MnTe$_2$ film was obtained under conditions of low growth temperature and high Te/Mn flux ratio, as confirmed by XRD analysis against the standard PDF card (Fig. S4).

Based on the XRD results, a phase diagram correlating flux ratio with growth temperature was successfully constructed (Fig. 2e) using the relative intensity ratio of the characteristic α-MnTe (0002) and γ-MnTe (111) diffraction peaks. The diagram clearly delineates the formation regions of each phase: α-MnTe dominates under high-temperature and high-flux-ratio conditions, while γ-MnTe is stabilized at lower temperatures and lower flux ratios. In contrast, the MnTe$_2$ phase forms preferentially in the low-temperature, high-flux-ratio region. These findings establish a reliable experimental guideline for the epitaxial growth of high-quality, pure-phase α-MnTe thin films.

To examine whether in-plane strain is introduced in the α-MnTe thin films during the growth process, we performed reciprocal space mapping (RSM) around the InP (222) and (531) peaks on a 30-nm-thick α-MnTe film at 300 K (Fig. 1f, g). The derived lattice parameters ($a$ = 4.19 Å, $c$ = 6.68 Å) indicate an in-plane tensile strain of ~1%, given the bulk α-MnTe value of 4.148 Å. Since the in-plane lattice constant of the InP(111) substrate (4.15 Å) is nearly identical, epitaxial mismatch is ruled out as the primary cause. The strain is instead attributed to the substantial difference in thermal expansion coefficients (CTE). The CTE of α-MnTe (1.6 × 10$^{-5}$ K$^{-1}$) [33] far exceeds that of InP (4.7 × 10$^{-6}$ K$^{-1}$) [34]. Therefore, during cooling from the growth temperature, the greater contraction of α-MnTe relative to InP results in the observed tensile strain and lattice expansion in the film. Importantly, the introduction of strain into the altermagnetic α-MnTe film can alter its magnetic structure, thereby affecting magnetostructural coupling [35]. Ultimately, strain in α-MnTe can modify the electronic structure and induce a giant spin-splitting effect [36, 37].

Raman spectra for both the α-MnTe and γ-MnTe films were obtained, with results shown in Fig. 3a and 3b. The two peaks located at 122 and 140 cm$^{-1}$ correspond to the A$_{1g}$ and E$_{TO}$ phonon modes, respectively, which originate from Te–Te bonds resulting from trace amounts of Te precipitated on the film surface [21, 38]. A distinct peak observed at 268 cm$^{-1}$ (see the enlarged view) is attributed to two-magnon (2M) scattering within the MnTe films [38]. The observed 2M phonon mode, arising from double spin-flip processes in the ground state [38], constitutes an unconventional scattering mechanism and can be also regarded as a characteristic feature of altermagnets. Notably, the Raman intensity of the A$_{1g}$ mode in γ-MnTe is nearly ten times stronger than that in α-MnTe (Figs. 3a,b and Figs. S5a,c). Based on this feature, we extracted the A$_{1g}$ Raman intensities of MnTe films grown under different Te/Mn flux ratios and growth temperatures, and constructed a flux-ratio-versus-growth-temperature phase diagram for MnTe films (Figs. S5e). Interestingly, this phase diagram agrees well with the one derived from XRD measurements (Fig. 2e), further confirming the reliability of the phase diagram in Fig. 2e.

The surface composition of the MBE-grown α-MnTe films was examined by X-ray photoelectron spectroscopy (XPS). Deconvolution of the Mn 2p core-level spectrum (Fig. 3c) reveals the spin-orbit components Mn 2p$_{3/2}$ and Mn 2p$_{1/2}$ at 641.6 eV and 653.3 eV, respectively, along with characteristic satellite peaks at 646.1 eV and 657.8 eV. Similarly, the Te 3d spectrum displays the Te 3d$_{5/2}$ and Te 3d$_{3/2}$ states at 572.2 eV and 582.6 eV, respectively (Fig. 3d). The higher-binding-energy doublet at 575.4 eV and 585.7 eV is assigned to Te–O bonds in a native surface oxide (TeO$_x$), consistent with previous studies on α-MnTe films [20, 22, 39].

To verify the crystal quality of the α-MnTe films, high-resolution atomic-scale imaging was performed using scanning transmission electron microscopy (STEM). A cross-sectional STEM specimen was prepared by focused ion beam (FIB), as shown in Fig. 4a. The film thickness of approximately 30 nm was clearly determined from the energy-dispersive X-ray spectroscopy (EDS) mapping images (Fig. 4b–e), which distinctly identified the InP substrate and the α-MnTe film. Fig. 4f shows an enlarged view of the region marked by the red rectangle in Fig. 4a, revealing a sharp interface between the α-MnTe film and the InP substrate. This indicates that the α-MnTe film is directly epitaxially grown on the InP substrate, rather than through a buffer layer as previously reported. A high-resolution atomic-scale image, taken from the region marked by the red dashed rectangle in Fig. 4f, is displayed in Fig. 4g. The atoms are arranged in a layer-by-layer manner, which aligns well with the theoretical structure (shown schematically at the right side of Fig. 4g). Fig. 4h presents the elemental profile of Mn and Te atoms in the α-MnTe film, extracted from the region marked by the red rectangle in Fig. 4g. The profile indicates an ordered arrangement of Mn and Te atoms. These results collectively confirm the successful epitaxial growth of high-quality α-MnTe thin films, thereby establishing a robust foundation for subsequent investigations into their transport properties.

Based on the aforementioned results, high-quality altermagnetic α-MnTe films were successfully grown by MBE. The next step is to perform transport measurements on these films to reveal the proposed altermagnetic properties. Here, we focus on anomalous Hall effect of α-MnTe film, with the corresponding data presented in Fig. 5. Fig. 5a shows the temperature-dependent resistance of a 30-nm-thick α-MnTe film. The inset illustrates the measurement configuration for the longitudinal resistance ($R_{xx}$) and anomalous Hall resistance ($R_{AHE}$). A Néel temperature ($T_N$) of approximately 300 K is clearly observed, as indicated by the red arrow. Below $T_N$, the α-MnTe film exhibits typical metallic behavior, consistent with previous reports [29]. Fig. 5b displays $R_{xx}$ (right axis) and $R_{AHE}$ (left axis) measured at 150 K, showing a clear rectangular hysteresis in $R_{AHE}$ in which the red arrows indicate the magnetic field sweeping from positive to negative, while the blue arrows indicate the reverse sweep. $R_{xx}$ also exhibits a corresponding non-monotonic trend, which together constitute a characteristic anomalous Hall signal.

The transverse Hall resistance $R_{xy}$ in a ferromagnetic conductor can be expressed by the empirical relation $R_{xy} = R_0 H + R_{\text{AHE}}$. Here, $R_0 H$ represents the ordinary Hall contribution ($\propto H$), and $R_{\text{AHE}}$ is the anomalous Hall resistance, which scales with the magnetization as $R_{\text{AHE}} \propto R_s M$, with $R_s$ being the material-dependent anomalous Hall coefficient. In general, the anomalous Hall effect is attributed to two distinct origins: extrinsic mechanisms and an intrinsic mechanism [40]. Extrinsic contributions, such as skew scattering and side jump, result from electron scattering by defects and impurities. In contrast, the intrinsic contribution arises from the Berry curvature of the material's electronic structures [41].

Fig. 5c and d display the temperature dependence of $R_{\text{AHE}}$ and $R_{xx}$ in the α-MnTe film, respectively, measured from 10 K to 250 K. Notably, an unexpected sign reversal of the anomalous Hall effect occurs around 75 K as the temperature increases from 10 K to 250 K. In the literature, a temperature-dependent sign reversal of the anomalous Hall effect has also been observed in ferromagnetic CrTe$_2$ films, which is attributed to changes in the Berry curvature [42]. Similarly, both the anomalous Hall effect and a sign reversal of its coefficient have also been achieved in bulk MnTe single crystals under external strain [43, 44]. The sign can be continuously tuned by varying the strain from compressive to tensile, and the reversal is attributed to the corresponding evolution of the Berry curvature in the electronic structure [45, 46]. In our case, extrinsic contributions can be excluded based on two lines of evidence. First, for an anomalous Hall effect dominated by defects or impurities, the sign of its coefficient is not expected to change with temperature. Second, STEM imaging reveals no discernible defects or impurities in the α-MnTe film. Magnetization measurements confirm that the α-MnTe film exhibits antiferromagnetic behavior, as shown in Supplementary Fig. S6.

Moreover, the in-plane strain in our α-MnTe films can be related to the anomalous Hall effect and its sign reversal. Recent theoretical studies show that strain-induced Fermi surface distortion can enhance the anomalous Hall conductivity by orders of magnitude [47]. Furthermore, applied strain substantially modifies transport signatures, including an order-of-magnitude enhancement in the planar Hall effect and anisotropic magnetoresistance [48].

Therefore, the observed anomalous Hall effect in the α-MnTe film is attributed to the Berry curvature of the material's electronic structure. The sign reversal of the anomalous Hall effect can be attributed to the strain in the α-MnTe film resulting from the CTE mismatch between the InP substrate and α-MnTe, with RSM measurements confirming a tensile strain of approximately 1% in the film.

## 3. Conclusion

In summary, we have successfully synthesized high-quality altermagnetic α-MnTe epitaxial films via systematic growth optimization. The work establishes a growth phase diagram, revealing the stability of the pure α-phase at high Te/Mn flux ratios and temperatures. Structural analysis confirms an atomically sharp interface and the presence of ~1% tensile strain originating from thermal expansion mismatch with the InP substrate. Raman spectroscopy further validates the phase stability, showing a characteristic weak signal for the α-phase. Crucially, transport measurements demonstrate a Néel temperature near room temperature (~300 K) and a pronounced anomalous Hall effect with a sign reversal around 75 K, attributed to Berry curvature modulation. These results collectively confirm the high crystalline quality and reveal the unique electronic properties of α-MnTe, underscoring its significant potential for scalable spintronic applications.

## 4. Experimental Section

**Growth of the α-MnTe Films**

The MnTe films were epitaxially grown on InP (111) substrates using a molecular beam epitaxy (MBE) system. Prior to growth, the InP (111) substrates were annealed at 500°C for 1 hour. After annealing, the substrates were cooled to the growth temperature. The chamber pressure during growth was maintained at $4 \times 10^{-10}$ Torr. High-purity Mn (99.999%) and Te (99.9999%) were evaporated from conventional effusion cells. During the growth of MnTe thin films, the temperature of the Mn source was fixed at 750°C, and the Te/Mn flux ratio was controlled by adjusting the temperature of the Te source. An *in situ* reflection high-energy electron diffraction (RHEED) system was employed to monitor the annealing of the InP (111) substrates and the growth of the MnTe films.

**Characterization of the α-MnTe Films**

*XRD and RSM Characterization.* High-resolution XRD and RSM measurements were performed at room temperature using a Rigaku SmartLab diffractometer with Cu K$_\alpha$ radiation ($\lambda$ = 1.5406 Å).

*Raman Characterization*. Raman measurements were conducted using a HORIBA LabRAM HR Evolution system equipped with a 532 nm non-polarized laser. The laser power was maintained at 8 mW to prevent sample degradation. *XPS Characterization.* The elemental composition and valence states were analyzed using an AXIS-ULTRA DLD-600W X-ray photoelectron spectroscopy system. *STEM Characterization.* The microstructure of the α-MnTe film was characterized using high-resolution scanning transmission electron microscopy (Thermo Fisher Titan Themis G2). The STEM specimen was prepared using an FEI Helios 600i focused ion beam (FIB) system.

**Transport Measurements**

Transport measurements were performed using a physical property measurement system (PPMS, DynaCool), where the temperature can be cooled down to 2 K and a magnetic field up to 14 T can be applied. A conventional six-probe device was prepared by attaching 25-µm diameter Au wires to the sample with Epotek H20E silver epoxy. An AC current was applied to the device using a Keithley 6221 current source. The voltage signals from the device were collected using an SR830 lock-in amplifier.

**Artificial Intelligence Generated Content (AIGC) Statement**

No artificial Intelligence Generated Content (AIGC) tools such as ChatGPT and others based on large language models (LLMs) are used in developing any portion of this manuscript.

## Acknowledgements

This work was supported by the National Key R&D Program of China (Grant No. 2022YFA1403700 and No. 2025YFA1411302), the National Natural Science Foundation of China (NSFC) (No. 12534003 and No. 12374111), the Guangdong Provincial Quantum Science Strategic Initiative (Grant No. GDZX2401002 and No. GDZX2301004), the Shenzhen Science and Technology Program (No. 20220814162144001), and the Joint Research Program between SUSTech and the National University of Singapore (NUS).

## Author contributions

T.-H. S. and X. D. contributed equally to this work. C. L. supervised the project. T.-H. S. and X. D. grew the thin films. W. H. performed the STEM measurements. M.-Y. Z. conducted the theoretical analysis and DFT calculations. Y. H. and C. W. performed the Raman measurements. L.-H. Y. assisted with material growth. J. Liu, M. Y., Q. G., and J. Li provided technological support for transport measurements. J.-J. S., T.-Y. X., X.-R. L., Y.-J. H., Y. D., X.-M. M., M. Z., and G. W. assisted with MBE system construction. T.-H. S., X. D., and C. L. wrote the manuscript with the help from all authors.

## Data Availability Statement

The data that support the findings of this study are available on request from the corresponding author. The data are not publicly available due to privacy or ethical restrictions.

## Competing Interests

The authors declare no competing interests.

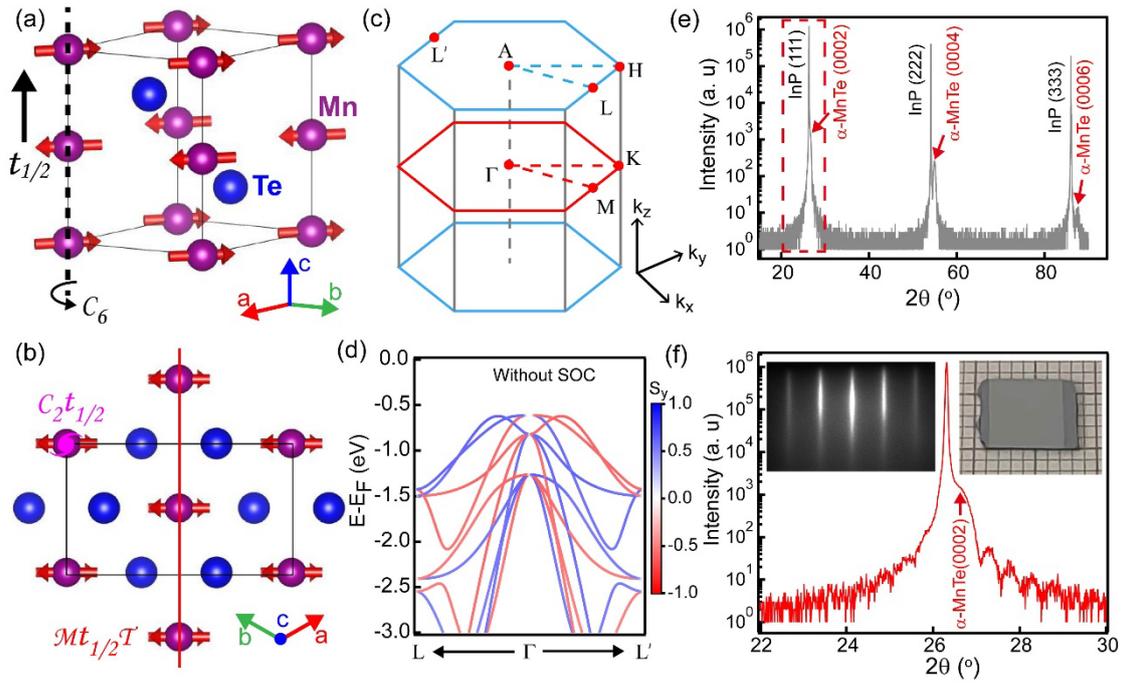

Figure 1. (a) Crystal structures of α-MnTe. (b) Projection of the crystal structure within the c-plane. The magnetic moments of Mn atoms are aligned within the c-plane, with adjacent atomic layers exhibiting antiparallel alignment, thereby forming an intralayer ferromagnetic structure and an interlayer antiferromagnetic coupling. (c) Schematic diagram of the first Brillouin zone and high-symmetry points of α-MnTe. (d) Spin-projected electronic structure calculated without SOC along the L–Γ–L′ path. The projected spin component ($S_y$) is defined to be along the direction of the magnetic moment in the real space, and parallel to Γ–M in the Brillouin zone. (e) The XRD results of the α-MnTe film. (f) Enlarged view of the α-MnTe (0002) diffraction peak for the region indicated by the red dashed rectangle in (e). The insets present the RHEED patterns (left) and an optical microscope image (right) of a macroscopic α-MnTe film against a millimeter grid. The film size reaches the centimeter scale.

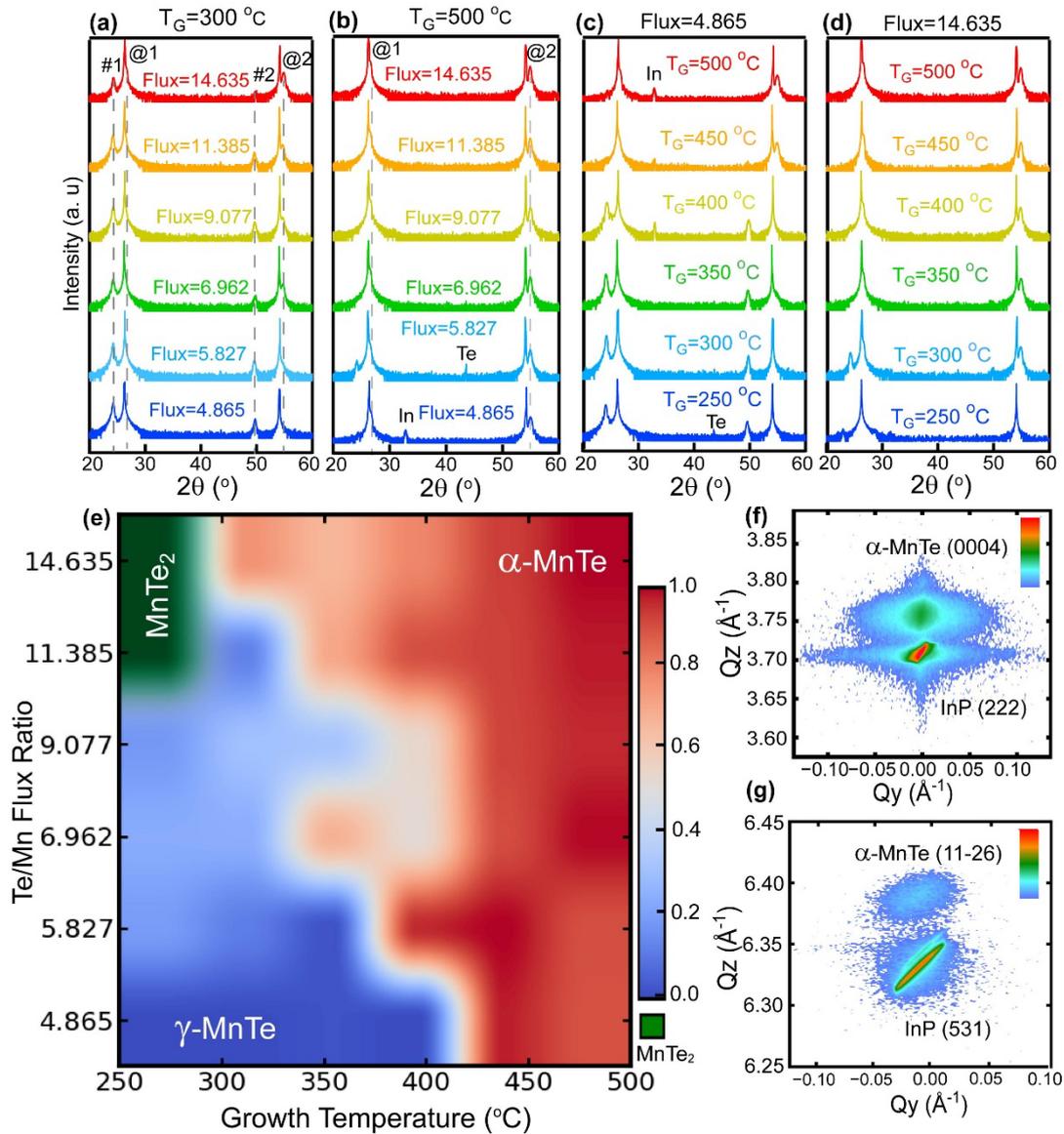

Figure 2. XRD phase analysis of MBE-grown MnTe films. (a, b) Te/Mn flux-ratio-dependent XRD scans at growth temperatures of 300°C and 500°C. @1, @2, #1 and #2 represent the α-MnTe (0002), α-MnTe (0004), γ-MnTe (111), and γ-MnTe (222) diffraction peaks, respectively. (c, d) Growth-temperature-dependent XRD scans at Te/Mn flux ratios of 4.865 and 14.635. (e) Growth-phase diagram plotted as Te/Mn flux ratio versus temperature. Colors indicate the intensity ratio between the α-MnTe (0002) (@1) and γ-MnTe (111) (#1) diffraction peaks. (f, h) Reciprocal space maps around the InP (222) and (531) reflections, from which the out-of-plane and in-plane lattice parameters were derived, respectively.

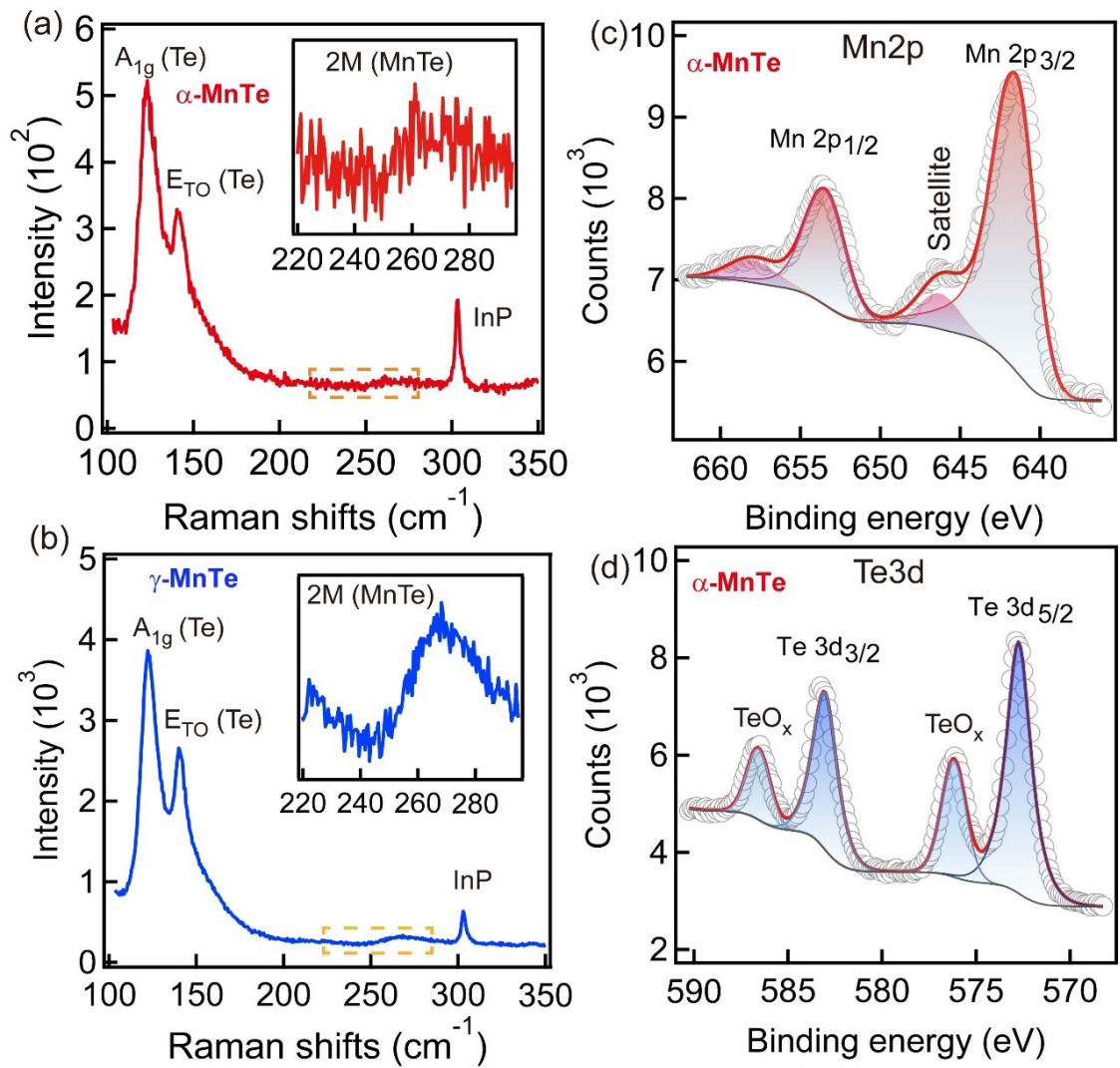

Figure 3. Raman and XPS analysis of α-MnTe and γ-MnTe. (a, b) Comparative Raman spectra (100 – 350 cm$^{-1}$) of α-MnTe and γ-MnTe, showing a tenfold intensity enhancement in the γ phase. Insets magnify the two-magnon scattering features. (c, d) XPS spectra of the Mn 2p and Te 3d core levels, respectively.

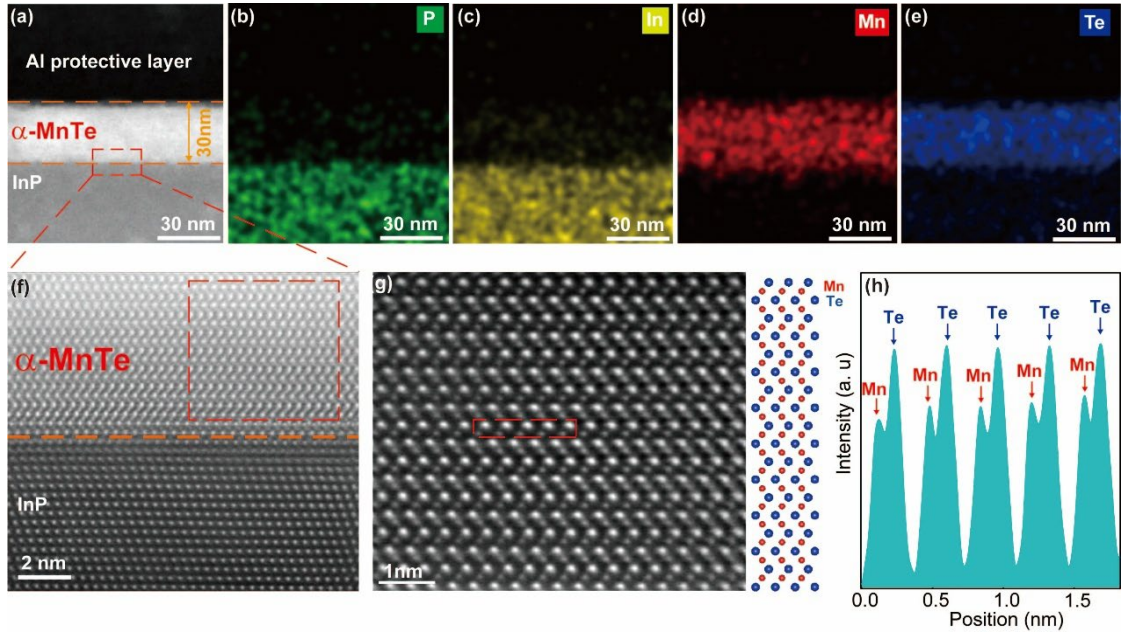

Figure 4. Microstructural and elemental characterization of α-MnTe films. (a) Cross-sectional schematic of the STEM specimen, showing an Al protective layer, a ~30-nm-thick α-MnTe film, and the InP substrate. (b–e) Corresponding energy-dispersive X-ray spectroscopy (EDS) elemental maps for P, In, Mn, and Te, respectively. (f) High-resolution STEM image of the interface region marked in (a), confirming an atomically sharp α-MnTe/InP interface. (g) Atomic-resolution image showing the layer-by-layer epitaxy of α-MnTe film, with the theoretical crystal structure model shown on the right. (h) Intensity profile derived from integrating the atomic column contrast within the red box in (g).

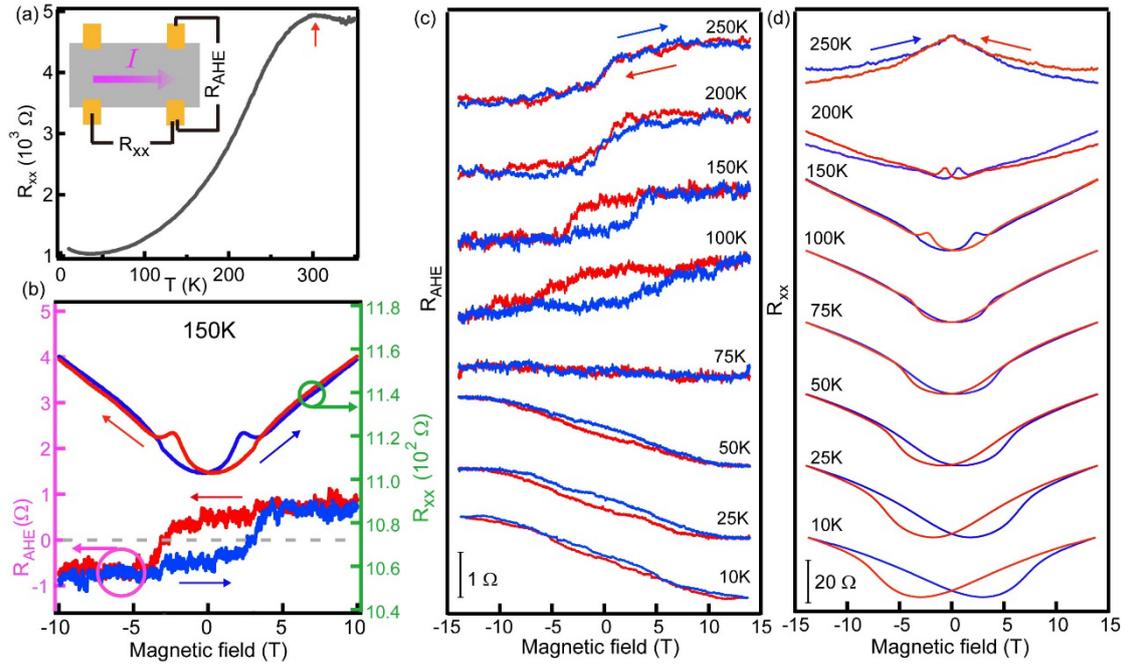

Figure 5. Electrical and anomalous Hall transport in α-MnTe films. (a) Temperature dependence of the longitudinal resistance ($R_{xx}$) from 5 K to 300 K. Inset: schematic of the measurement geometry for $R_{xx}$ and the anomalous Hall resistance ($R_{AHE}$). (b) The anomalous Hall resistance ($R_{AHE}$, left axis) and $R_{xx}$ (right axis) as a function of magnetic field at 150 K. Red (blue) arrows (same below): magnetic field sweeping from positive to negative (from negative to positive). (c) Temperature-dependent anomalous Hall resistance ($R_{AHE}$) from 10 K to 250 K. A sign reversal of the anomalous Hall effect is seen to occur around 75 K. (d) Temperature dependence of $R_{xx}$ over the same temperature range.